\documentclass[preprint,showpacs,preprintnumbers,amsmath,amssymb]{revtex4}

\usepackage{graphicx}
\usepackage{dcolumn}
\usepackage{bm}

\newcommand{\wzero}{$^{180}\rm{W}~$}
\newcommand{\wone}{$^{181}\rm{W}~$}

\newcommand{\wfour}{$^{184}\rm{W}~$}
\newcommand{\wfive}{$^{185}\rm{W}~$}
\newcommand{\wsix}{$^{186}\rm{W}~$}
\newcommand{\wseven}{$^{187}\rm{W}~$}
\newcommand{\wzeron}{$^{180}\rm{W}$}
\newcommand{\wonen}{$^{181}\rm{W}$}

\newcommand{\wfourn}{$^{184}\rm{W}$}
\newcommand{\wfiven}{$^{185}\rm{W}$}
\newcommand{\wsixn}{$^{186}\rm{W}$}
\newcommand{\wsevenn}{$^{187}\rm{W}$}
\newcommand{\crzero}{$^{50}\rm{Cr}~$}
\newcommand{\crone}{$^{51}\rm{Cr}~$}

\newcommand{\ngamma}{$(n,\gamma)~$}
\newcommand{\ngamman}{$(n,\gamma)$}
\newcommand{\er}{$\pm~$}
\newcommand{\gm}{$\gamma s~$}

\begin{document}


\title{Measurement on the thermal neutron capture cross section of \wzero}

\author{W.G. Kang}
\author{Y.D. Kim}
\email{ydkim@sejong.ac.kr}
\author{J.I. Lee}
\affiliation{Department of Physics, Sejong University, Seoul, 143-747, Korea}

\author{I. S. Hahn}
\author{A.R. Kim}
\affiliation{Department of Science Education, Ewha Woman's University, Seoul 120-750, Korea}

\author{H. J. Kim}
\affiliation{Physics Department, Kyungpook National University, Daegu, 702-701, Korea}

\date{\today}

\begin{abstract}
We have measured the thermal neutron capture cross section for \wzero nucleus. There is only one previous data on
this cross section with a value of 30 $^{+300\%}_{-100\%}$ barn. 
To consider \wone as a low energy neutrino source,
the thermal neutron capture cross section should be measured more precisely to estimate the production rate 
of \wone inside a nuclear reactor. 
We measured the cross section of \wzero with a natural tungsten foil and obtained a new value of 21.9 \er 2.5 barn.
\end{abstract}



\maketitle

\section{Motivation}

Neutrino oscillation experiments such as Super-Kamiokande (SK) \cite{Ashie:2004mr}, K2K \cite{Ahn:2002up}, 
SNO \cite{Ahmed:2003kj}, and the KAMLAND \cite{Eguchi:2003gg}, have achieved great progresses 
in understanding the masses and mixing angles of the neutrinos.
The non-zero masses and the large mixing angles of the neutrinos are confirmed and stimulate
further theoretical models on the neutrino masses.
The above neutrino oscillation experiments detected the neutrinos from the sun (SK, SNO), from 
accelerator (K2K), and from nuclear reactors (KAMLAND, CHOOZ). Therefore the neutrino sources are 
fixed in their locations, which limits the flexibility of the experiments to some extent.
In addition to these neutrino sources, artificial neutrino source (ANS) has been studied for the 
calibration of neutrino detectors such as
GALLEX \cite{Anselmann:1994ar} and SAGE \cite{Abdurashitov:1996dp}.
The ANS refers the neutrinos from a beta decaying nuclei. For example, the neutrinos from electron capture of 
\crone nuclei are mostly mono-energetic at the energy of about 750 keV.
About 1 MCi(million Curie) of \crone was produced 
by \ngamma capture reaction inside a nuclear reactor with enriched  \crzero to calibrate the solar 
neutrino detector \cite{Abdurashitov:1996dp,Anselmann:1994ar}.

Besides the calibration of a neutrino detector, the radioisotope neutrino 
source can be used to 
study non-standard neutrino properties and possibly for neutrino oscillation experiment. 
For example, about 1 MCi neutrino source can be used to measure the magnetic 
moment of the neutrinos with a neutrino detector closely located to the source. 
Recently two groups, TEXONO \cite{wong:2006nx} and 
MUNU \cite{Daraktchieva:2005kn}, reported most stringent upper limits on the neutrino magnetic moment as 
$\mu_{\nu} < 7.4 \times 10^{-11} \mu_{B}$ and $\mu_{\nu} < 9.0 \times 10^{-11} \mu_{B}$ respectively from the
measurements of $\nu e$ elastic scattering energy spectra with the reactor neutrinos.
The neutrino magnetic moment is one of the most fundamental properties of neutrinos and it is anticipated 
that one can improve the sensitivity if a strong neutrino source 
with a proper radioisotope is available.

There are other candidates for a neutrino source besides \crone, such as \wonen, $^{170}$Tm, and $^{147}$Pm, 
etc \cite{Barabanov:1997fv}.
An appropriate radioisotope for ANS should have characteristics such as; (1) relatively long decay time 
between 10 days and 10 year to 
perform an experiment, (2) low gamma intensities for safety issue, (3) abundances of mother nuclei should be 
relatively large for low cost, (4) thermal neutron capture cross section should be large.
Among the nuclei, $^{147}$Pm is produced from nuclear spent fuel, and the others are produced by \ngamma 
reaction in nuclear reactors. \wone has good properties except the low abundance of \wzero at the level of 0.12\%.
Therefore an enrichment is necessary for tungsten, and the production cost depends on the amount of material  
needed to produce 
the desirable activity, which is usually more than 1MCi. In this respect, the thermal neutron capture cross section of \wzero is an important 
parameter to know. 
Until now, however, there
is only one data on the thermal neutron capture cross section on this nucleus, 30 $^{+ 300 \%}_{-100 \%}$ barn, 
measured by Pomerance more than 50 years ago \cite{pomerance:1952}. Since the uncertainty of this data is very large,
it is necessary to measure the cross section more
precisely to determine if \wone is a good candidate for ANS.

To measure the capture cross section, we irradiated
natural tungsten foils in a thermal neutron irradiation facility and 
measured the \gm from \wone decay. In the natural tungsten foil, \wfour and \wsix are more abundant, and the capture
cross sections were measured with sufficiently small uncertainties. Therefore, we can obtain the \wzero capture
cross section by a comparison with the activity measurements of \wfive and \wsevenn. In this report we describe 
the measurement of a capture cross 
section of \wzeron\ngamman\wone reaction with a much smaller error than the previous measurement.

\section{Experimental Setup}
An irradiation area at HANARO research reactor facility in Korea
was used for the measurement. At HANARO, a neutron irradiation facility for 
BNCT (Boron Neutron Capture Therapy) consists of a water shutter, a fast neutron and gamma
ray filter, a liquid nitrogen cooling system, a beam collimator, and shieldings \cite{mskim05}.
The cadminium ratio (Cd ratio) at this facility is known to be larger than 100 at the position of irradiation. 
We prepared two identical tungsten foils (99.9\% pure) with a size of 50.1mm X 50.1mm X 0.138mm (6.644g) 
to measure the Cd ratio at the same time of the irradiation.
The thickness of the foil was calculated
by dividing the mass of the foil by the area with a density of 19.25$g/cm^{3}$. One foil was sandwiched by two 
Cadminium foils 
of about 1mm thickness. The thermal neutron flux was previously measured at this beam 
line as about 7-8$\times$$10^{8}$ neutrons/sec/cm$^{2}$. 

Table \ref{tungstenisotopes} shows the informations of stable isotopes in natural tungsten foil \cite{toi}.
The idea is that we can obtain the capture cross section of \wzero with respect to the capture cross
sections of \wfour and \wsix which were measured with much smaller errors. This method has the advantage 
of cancellation 
of potential systematic errors from thermal
neutron flux, flux profile, foil thickness, foil size, and irradiation time etc. Neither the absorption effect
inside the tungsten foil contributes to the errors in the final cross section since 
all the stable tungsten
isotopes will see the same amount of the neutron flux regardless of the absorption.
As shown in the table \ref{tungstenisotopes}, there are \gm with sufficiently long half-lives for \wfiven, 
\wsevenn, and \wonen, from which we can obtain the production rates of the three radioisotopes. There are
two more \gm from \wseven decay with significant gamma intensities, 134.2 keV and 479.5 keV, but these
gammas was not used in current analysis. 134.2 keV gamma is not separable from 136.3 keV gamma from \wone 
and 479.5 keV gamma was overlapped with other unidentified gammas.

\begin{table}[h]
\caption{Stable isotopes in natural tungsten. $T_{1/2}$, E$_{\gamma}$, and I$_{\gamma}$(gamma intensity) are for the
the produced radioisotopes by \ngamma reaction. The values in the parentheses in $I_{\gamma}$ indicates
the associated errors.}
\label{tungstenisotopes}
\begin{ruledtabular}
\begin{tabular}{ccccccc}
Isotope              & Abun.     & $\sigma$  & $T_{1/2}$ & E$_{\gamma}$ & I$_{\gamma}$  \\
                     & (\%)      & (barn)    & (day)     & (keV)        &   (\%)   \\
\hline
\wzeron              &  0.12     & 30 $^{+90}_{-30}$ \footnotemark[1]    &  121.2    &  136.3       & 0.0311(10) \\
                     &           & 21.9 \er 2.5\footnotemark[2]   &  &  152.3       & 0.083(3) \footnotemark[5]   \\ \hline
\wfourn  & 30.6      & 1.76 \er 0.09 \footnotemark[3]  &  75.1     &  125.4       & 0.0192(3)   \\ \hline
\wsixn   &  28.4     & 39.5 \er 2.3 \footnotemark[4]   &  0.988   &  551.5       & 5.08(17)   \\
         &           &           &           &  618.4       & 6.28(21)   \\
         &           &           &           &  685.8       & 27.3(9)   \\
         &           &           &           &  772.9       & 4.12(13)   \\
\end{tabular}
\end{ruledtabular}
\footnotetext[1] {Reference \cite{pomerance:1952}.}
\footnotetext[2] {This work.}
\footnotetext[3] {Reference \cite{bondarenko05}.}
\footnotetext[4] {Reference \cite{Karadag04}.}
\footnotetext[5]{The gamma intensity of this level is written as 0.0083 by mistake in all the existing 
database including NNDC database, Table of Isotopes (8th Ed.), Nuclear Data Sheets \cite{SCWu:2005ab}, and NUDAT.
The correct value, 0.083, was reported in the paper of \cite{indira79}. We reported this mistake to NNDC database
group, and it is corrected in NNDC database at the moment of this writing.}
\end{table}

We irradiated both the enclosed and open foils for 5 hours at the BNCT facility, 
and the irradiated foils are left for 12 days or more to reduce 
the activity of the foils. Even though the half-life of \wseven is only 1 day, we still have enough 
counts from the decay of \wsevenn.
The irradiated tungsten foils were measured with a low background HPGe detector located at 700 meter 
deep underground at Yangyang laboratory of Dark 
Matter Research Center (DMRC) in Korea. The relative efficiency of the HPGe detector is 100\%, and the detection 
efficiencies are measured with absolute intensity sources before, and compared with GEANT4 simulations \cite{lee:2007iq}.

\section{Results}
\begin{figure}
\includegraphics[width=9cm]{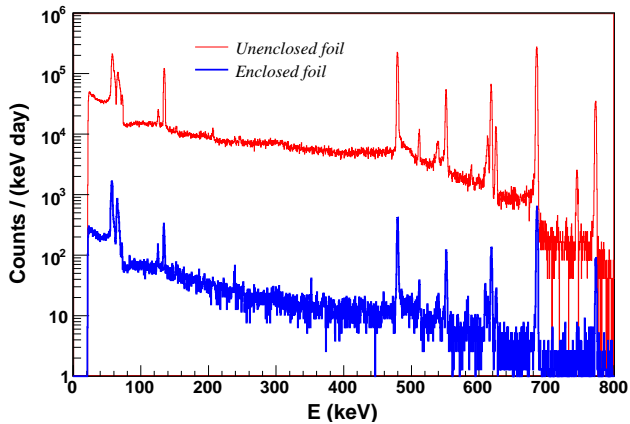}
\caption{\label{cdratio} HPGe spectrum of tungsten foils with (lower) and without (upper) Cd enclosure.}
\end{figure}

Figure \ref{cdratio} shows the HPGe spectra  of two tungsten foils with and without cadminium enclosure. The Cd ratio has been 
estimated by the ratio of the \gm from \wseven peaks, which are dominant in these spectra. The Cd ratio was 
obtained to be 245\er 10, which guarantees the activities we obtain from the open foil has little contribution
from non-thermal neutrons.

\begin{figure}
\includegraphics[width=9cm]{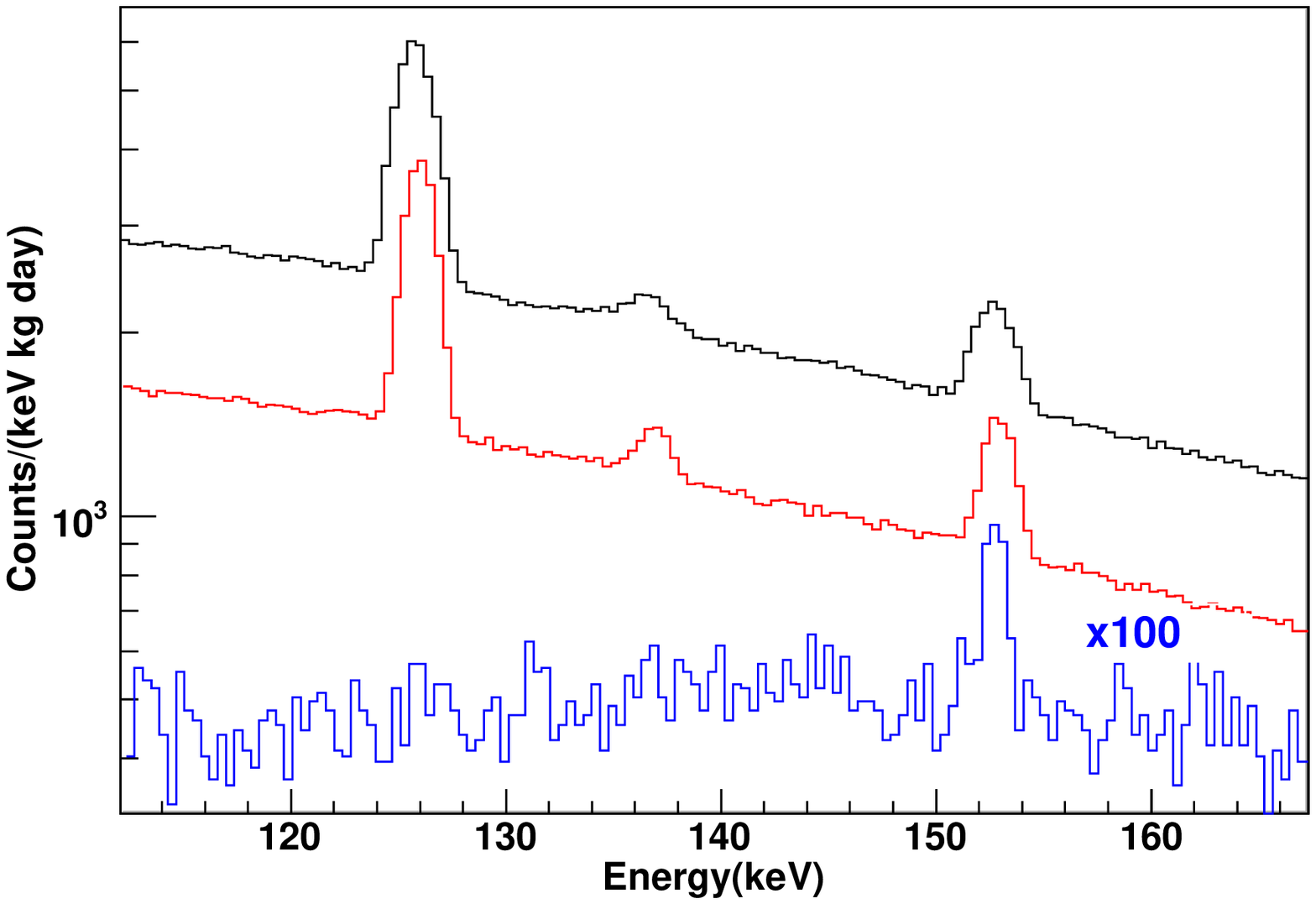}
\includegraphics[width=9cm]{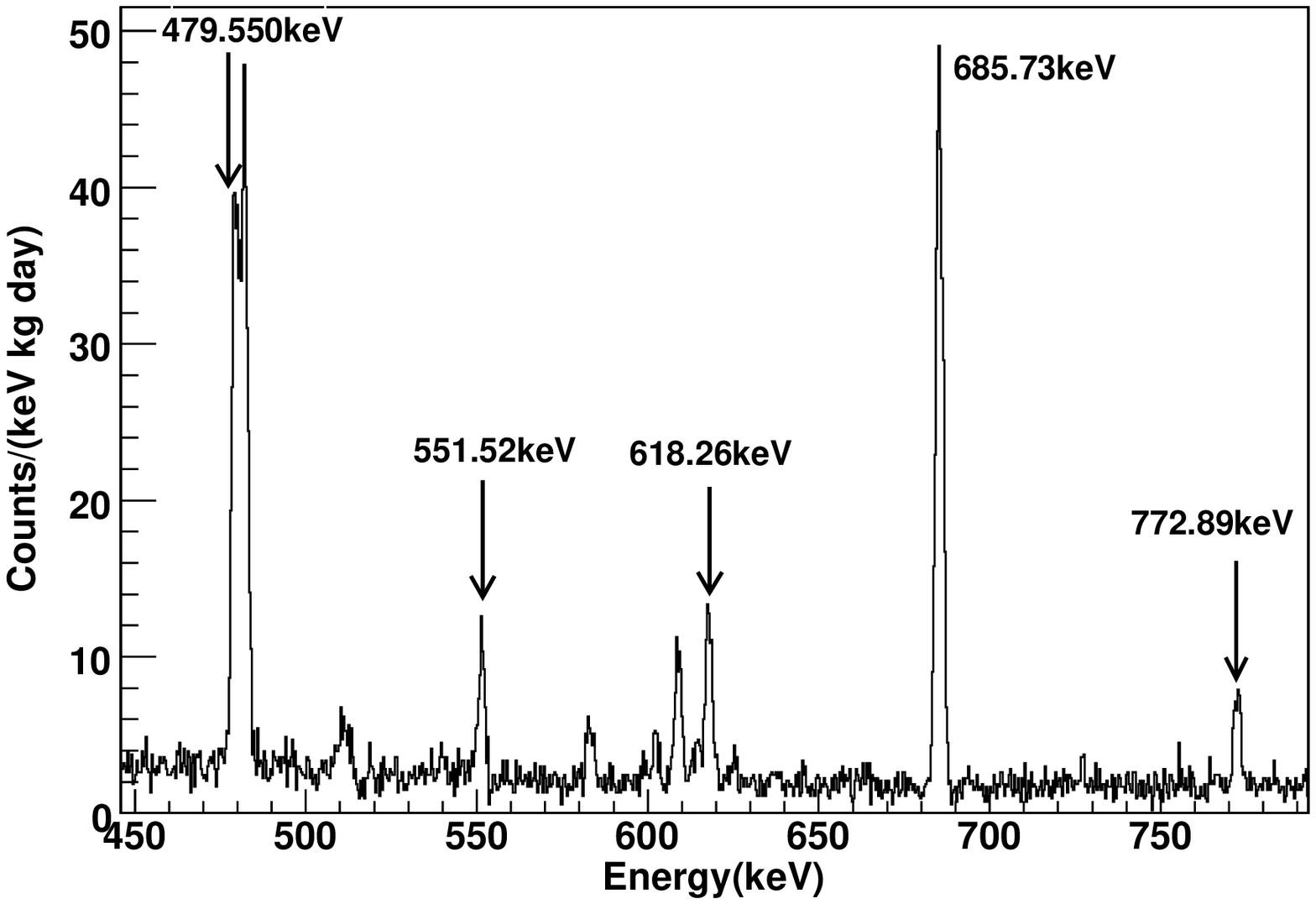}
\caption{\label{peaks} TOP : Energy spectra of the tungsten foil in the energy region of \wone and \wfive gamma 
peaks obtained after 25 days (upper), 81 days (middle), 993 days (lower, multiplied by 100 for easy comparison) 
since the end of irradiation period. 
BOTTOM : same as TOP in the energy region of \wseven obtained after 25 days.}
\end{figure}

Figure \ref{peaks}:TOP shows the measured HPGe spectra of gamma peaks at the energies of 136.3, 152.3 keV (\wonen), 
and 125.4 keV (\wfiven). The upper spectrum was obtained with a measurement of 5.4 days starting 
25.0 days after the irradiation.
Even after 25 days, there are continuous high background events below 400 keV. This was due to the bremsstrahlung 
photons from the beta decay of \wfive to the ground state of $^{185}Re$ (branching ratio 99.93\%), 
which has a Q value of 433 keV.
Even with the bremsstrahlung background, we 
can still observe 125.4, 136.3, and 152.3 keV gamma peaks clearly. We also show the spectra obtained at 
81 days (middle spectrum) and 993 days (lower spectrum)
since the end of the irradiation period. The data obtained at 81 days after the irradiation was used to get the
counts of the 136.3 keV peak of \wone since the 25 days data still has about 10 \% counts from 134.2 keV peak of
\wsevenn. We also confirmed that the half-lives of these peaks are consistent with the 
expectation of \wone and \wfiven.
The bremsstrahlung background was reduced significantly in the spectrum of last data.

\begin{figure}
\includegraphics[width=9cm]{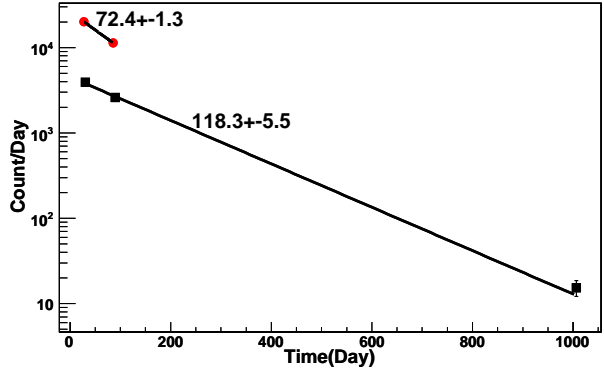}
\caption{\label{halflife} The activity of 125.4 keV (upper) and 152.3 keV (lower) \gm as a function of time of measurements. 
The half-lives of \wone and \wfive obtained from the data are written in the figure in units of days.}
\end{figure}

Figure \ref{peaks}:BOTTOM shows the peaks of 
551.5, 618.4, 685.8, and 772.9 keV (\wsevenn) in the higher energy region of the upper spectrum of 
Figure \ref{peaks}:TOP.
The 479.5 keV peak is overlapped with another peak and is not used in the analysis.
Figure \ref{halflife} shows the count rates of 152.3 keV and 125.4 keV \gm obtained at various measurement times after the
irradiation. We obtained the half-lives of \wfive and \wseven 
as 72.4 \er 1.3 day and 118.3 \er 5.5 days respectively. The half-life of \wfive is slightly off from the value in the
database.

\begin{table}[h]
\caption{The measured net counts of the \gm from irradiated tungsten sample with HPGe detector. Data are for
a measurement after 25 days from irradiation. $\eta$ is detection efficiency and R is production rate.}
\label{measurements}
\begin{ruledtabular}
\begin{tabular}{ccccccc}
             & $E_{\gamma}$  & Counts           & $\eta$     & R  & R$_{av}$ \\
             & (keV)         &                  & (\%)        & \multicolumn{2}{c}{$\times 10^{5}/sec$} \\
\hline
\wonen        &  136.3        & 7644 \er 614   & 12.2        & 4.27 \er 0.92 & 4.14 \er 0.62\\ 
             &  152.3        & 21208 \er 650  & 13.4        & 4.04 \er 0.83 &                \\ \hline
\wfiven       &  125.4        & 111357 \er 984 & 10.9        & 77.0 \er 15.5 & 77.0 \er 15.5 \\ \hline
\wsevenn      &  551.5        & 330 \er 33     & 8.5         & 1942 \er 439 & 1993 \er 221\\
             &  618.4        & 388 \er 36     & 8.0         & 1967 \er 439 &   \\
             &  685.8        & 1615 \er 60    & 7.5         & 2002 \er 413 & \\
             &  772.9        & 235 \er 28     & 7.0         & 2077 \er 488 & \\
\end{tabular}
\end{ruledtabular}
\end{table}

Table \ref{measurements} shows the measured count rates and errors.
The production rate of radioisotopes from \ngamma reaction during the irradiation can be written as 
\begin{equation}
R = F \frac{m \alpha A_{0}}{w} \sigma .
\end{equation}
Here $F$ is the neutron flux, m is the total mass of the foil, $\alpha$, $w$, and $\sigma$ are the capture cross section, 
mass number, and the relative abundance of target isotope, respectively. $A_{0}$ is the Avogadro's number.
The thermal neutron capture cross section of \wzero can be obtained with the existing capture cross section 
data of \wfour \cite{bondarenko05} and \wsix \cite{Karadag04} as
\begin{equation}
\sigma_{180} = \frac {\alpha_{i} w_{180} R_{180}} {\alpha_{180} w_{i} R_{i}} \sigma_{i}.
\label{sigma}
\end{equation}
Here the index $i$ refers to \wfour or \wsixn.
The production rate R is calculated from the data as,
\begin{equation}
R_{i}^{j} =\frac {C_{i}^{j}} {\tau_{i}~ I_{i}^{j}~ \eta_{i}^{j}~ (1-e^{-T/\tau_{i}})
(e^{-t_{1}/\tau_{i}} - e^{-t_{2}/\tau_{i}})}.
\label{rate}
\end{equation}
Here  $C_{i}^{j}, ~I_{i}^{j}, ~\eta_{i}^{j}$ are the net count, the gamma
intensity, and the detection efficiency of {\it j~}th gamma from {\it i~}th isotope respectively, and 
$R_{i}^{j}$ is the production rate calculated with the net gamma count.
$\tau_{i}$ is the mean decay time of the produced radioisotope, T is the irradiation time, 
and $t_{1}, t_{2}$ are the starting and stop time of the HPGe measurement since the end of the irradiation period.

The production rates of each radioisotope are in the last two columns of 
Table \ref{measurements}.  The errors in the production rates are due to the errors in count statistics,
gamma intensity \cite{SCWu:2005aa,SCWu:2005ab}, and detection efficiency. 
Unfortunately, the efficiencies 
of the HPGe detector at underground is not accurately measured since we observed discrepancies between calibrated source 
measurements and the GEANT4 simulations. The conservative estimation of the discrepancies are in the order of 20\% for 
the energy region over 100 keV, and it contributes most for the uncertainties
in the production rates. The errors of the abundances of each isotopes are negligible compared 
with the other errors, so neglected in the analysis.
In the case of \wone and \wsevenn, the production rates obtained with multiple \gm are consistent with each other 
as expected.
The weighted mean of the production rates are in the last column of Table \ref{measurements}.
Though, the thermal neutron flux cancels out in the cross section calculation, we can calculate the thermal 
neutron flux with the known cross section of \wfive and \wsevenn. The obtained flux results 
in 7.6 \er 0.7 $\times 10^{8}/cm^{2}/sec$, which is consistent with the value reported for the BNCT facility previously.

The capture cross section of \wzero is calculated using equation \ref{sigma} with the known 
cross section of 1.76 \er 0.09 barn for \wfour and 39.5 \er 2.3 barn for \wsixn, and we obtained 
24.7 \er 3.9 barn and 20.1 \er 3.2 barn
respectively. The two cross sections are consistent within the uncertainties. The final \wzero capture cross 
section is obtained
with the weighted mean of the two values, and it is 21.9 \er 2.5 barn.
While our new cross section is consistent with the data by Pomerance \cite{pomerance:1952}, the small 
uncertainty makes it possible
to evaluate the feasibility of \wone as a neutrino source, which was too ambiguous due to the large error in the 
Pomerance data.

\section{Discussion}

Though the present result can be improved with a better understanding of the HPGe detector efficiency in the future,
it is accurate enough to evaluate the feasibility of \wone as a neutrino source more realistically. 
Considering that \wone enrichment is technically difficult, \wone seem to be not very attractive. Instead, 
$^{170}$Tm would be a better candidate in the respect of the cost, but one has to figure out the safety issue 
related the high gamma-ray flux, and bremsstrahlung photons.
The measurement method we used in this report can be utilized for other low abundance isotopes.
In summary, we measured the thermal neutron capture cross section of \wzero as 21.9 \er 2.5 barn.
Though \wone is a good candidate for artificial neutrino source, the production of \wone would be too expensive
considering the thermal neutron capture cross section measured in this work.

\begin{acknowledgments}
This work was supported by the Korea Research Foundation (Grant KRF-2002-070-C00027). The authors thank
Dr. M.S. Kim and 
the staffs of the BNCT facility of HANARO research reactor for the neutron irradiation 
and the DMRC group for using the HPGe detector at underground.
\end{acknowledgments}

\bibliography{neutrinos}

\end{document}